\documentclass[12pt]{article}
\usepackage{graphicx}
\usepackage{subcaption}
\usepackage{cite}
\usepackage{lineno}

\newcommand\pubnumber{}
\newcommand\pubdate{\today}

\def\institute{Laborat\'orio de Instrumenta\c c\~ao e F\'isica de Part\'iculas (LIP)\\
Universidade do Minho, Braga, Portugal}
\def\support{\footnote{The author was funded by Funda\c c\~ao para a Ci\^encia e Tecnologia with the grant SFRH/BD/129321/2017 and projects CERN/FIS-PAR/0008/2017 and PTDC/FIS-PAR/29147/2017,
COMPETE2020-Portugal2020, POSI through project POCI-01-0145-FEDER-029147.
\\Copyright 2018 CERN for the benefit of the ATLAS Collaboration. Reproduction of this article or parts of it is allowed as specified in the CC-BY-4.0 license.}}

\def\Title#1{\begin{center} {\Large #1 } \end{center}}
\def\Author#1{\begin{center}{ \sc #1} \end{center}}
\def\Address#1{\begin{center}{ \it #1} \end{center}}

\newcommand\pubblock{\rightline{\begin{tabular}{l} \pubnumber\\
         \pubdate  \end{tabular}}}
\newenvironment{Abstract}{\begin{quotation}  }{\end{quotation}}
\newenvironment{Presented}{\begin{quotation} \begin{center} 
             PRESENTED AT\end{center}\bigskip 
      \begin{center}\begin{large}}{\end{large}\end{center} \end{quotation}}

\def\beq{\begin{equation}}
\def\eeq#1{\label{#1}\end{equation}}
\def\eeqn{\end{equation}}

\def\beqa{\begin{eqnarray}}
\def\eeqa#1{\label{#1}\end{eqnarray}}
\def\eeqan{\end{eqnarray}}

\let\bar=\overbar

\def\Dslash{\not{\hbox{\kern-4pt $D$}}}
\def\dslash{\not{\hbox{\kern-2pt $\del$}}}

\def\msb{{\bar{\ssstyle M \kern -1pt S}}}

\begin{document}
\begin{titlepage}
\pubblock

\vfill
	\Title{Search for flavour-changing neutral currents \textit{tZ} interactions in\\ \textit{pp} collisions at $\sqrt{s}$=13 TeV with ATLAS}
\vfill
\Author{Ana Peixoto\support \\On behalf of the ATLAS Collaboration}
\Address{\institute}
\vfill
\begin{Abstract}
A search for flavour-changing neutral currents (FCNC) processes in proton-proton ($pp$) collisions at a centre-of-mass energy of 13 TeV with the ATLAS detector at the CERN Large Hadron Collider (LHC) is presented. The analysed data collected during the years of 2015 and 2016 corresponds to an integrated luminosity of 36.1 fb$^{-1}$. A search considering top-quark pair-production events is performed, with one top-quark decaying through the dominant Standard Model (SM) mode $t\to Wb$, and the other through the $t\to qZ$ ($q$=$u,c$) FCNC channel. The data are consistent with the SM expectation and the observed and expected upper limits on the branching ratio of $t\to uZ$ and $t\to cZ$ are set at 95\% confidence level representing an improvement of about a factor 3 compared with the Run-1 data results from the ATLAS Collaboration.
\end{Abstract}
\vfill
\begin{Presented}
$11^\mathrm{th}$ International Workshop on Top Quark Physics\\
Bad Neuenahr, Germany, September 16--21, 2018
\end{Presented}
\vfill
\end{titlepage}
\def\thefootnote{\fnsymbol{footnote}}
\setcounter{footnote}{0}

\section{Introduction}

Discovered in 1995, the top-quark is the heaviest elementary particle in the SM and decays almost exclusively to a $W$ boson and a bottom-quark \cite{topmass}. Within the SM, the FCNC processes are forbidden at tree-level due to the Glashow-Iliopoulos-Maiani mechanism and suppressed at higher orders because of the unitary of the Cabibbo-Kobayashi-Maskawa matrix \cite{GIM}. On the other hand, there are several Beyond the Standard Model (BSM) scenarios with FCNC decays at tree-level with enhanced branching ratio predictions by several orders of magnitude \cite{Theory}. Therefore, any evidence of these processes would be an indication of new physics.

Experimental limits on the branching ratio of FCNC $t\to qZ$ were established by several experiments \cite{ALEPH,DELPHI,OPAL,L3,LEP,ZEUS,CDF,D0}. The most stringent limits before this analysis were set by the CMS Collaboration with $\mathcal{B}$($t\to uZ$) $<$ 2.4 $\times$ 10$^{-4}$ and $\mathcal{B}$($t\to cZ$) $<$ 4.5 $\times$ 10$^{-4}$ obtained at 95\% confidence level using data collected at $\sqrt{s}$ = 13 TeV \cite{CMS}.

The present analysis reports a search for the FCNC decay $t\to qZ$ in top-quark pair-production events with one top-quark decaying through the dominant SM mode (with a $W$ boson and a bottom-quark) and the other through the FCNC mode. Just leptonic decays from the $W$ and $Z$ bosons were considered. The final state of the signal process is then composed by three isolated charged leptons, at least two jets with exactly one being tagged as a jet containing $b$-hadrons, and missing transverse momentum from the neutrino.


\section{Event selection}

The analysis uses $pp$ collision data at a centre-of-mass energy of 13 TeV collected in the years 2015 and 2016 corresponding to an integrated luminosity of 36.1 fb$^{-1}$ with the ATLAS detector at the LHC \cite{detector}. The modelling of the signal and background processes is studied through Monte-Carlo (MC) simulation samples. The main source of background events containing three prompt leptons are $t\bar{t}Z$, $tZ$ and diboson production. The estimate of the contribution from these backgrounds is obtained from the MC simulation. Events with two or fewer real leptons and non-prompt leptons originate an additional background. The contribution from such background is estimated by means of a semi-data-driven method using the MC samples corrected by data in the specific control regions.

The signal process produces a final state characterised by exactly three leptons with $|\eta|$ $<$ 2.5 and $p_{\rm T}$ $>$ 15 GeV. The $Z$ boson is reconstructed from the two leptons that have the same flavour, opposite charge and a reconstructed mass within 15 GeV of the $Z$ boson mass ($m_Z$) \cite{topmass}. In case of two $Z$ boson candidates, the one with the reconstructed mass closest to $m_Z$ is selected as the $Z$ boson candidate. Additionally, the events with $E^{\rm miss}_{\rm T}$ $>$ 20 GeV and at least two jets, where one must be $b$-tagged, are selected. All jets must have $p_{\rm T}$ $>$ 25 GeV and
$|\eta|$ $<$ 2.5.

The reconstruction of the top-quark candidates from the distinct decay particles is achieved using a $\chi^2$ minimisation method based on energy-momentum conservation. The method consists in minimising the following expression without constraints: 
\begin{equation}
 \chi^2 = \frac{(m_{j_{a}l_{a}l_{b}}^{reco} - m_{t_{FCNC}})^2}{\sigma_{FCNC}^2} + \frac{(m_{j_{b}l_{c}\nu}^{reco} - m_{t_{SM}})^2}{\sigma_{SM}^2} + \frac{(m_{l_{c}\nu}^{reco} - m_{W})^2}{\sigma_{W}^2} \nonumber
\end{equation}
 where $m_{j_{a}l_{a}l_{b}}^{reco}$, $m_{j_{b}l_{c}\nu}^{reco}$ and $m_{l_{c}\nu}^{reco}$ are the reconstructed masses of the $qZ$, $bW$ and $l\nu$ systems, respectively. The $m_{t_{FCNC}}$, $m_{t_{SM}}$ and $m_{W}$ parameters are the central values of the masses of the reconstructed top-quarks and $W$ boson with the correspondent widths ($\sigma_{FCNC}$, $\sigma_{SM}$ and $\sigma_{W}$) obtained through simulated signal events. For each jet combination, $j_a$ corresponds to any jet while $j_b$ is assigned to the $b$-tagged jet. The transverse momentum of the neutrino from the $W$ boson decay is set to the value of $E^{\rm miss}_{\rm T}$. The minimisation of the $\chi^2$ equation provides the longitudinal component of the neutrino momentum ($p^{\nu}_{z}$). After taking into account all combinations, the one with the minimum $\chi^2$ is chosen. 
 
 The signal region selection is complete with the requirements of $|m_{j_{a}l_{a}l_{b}}^{reco}$ - 172.5~GeV$|$ $<$ 40~GeV, $|m_{j_{b}l_{c}\nu}^{reco}$ - 172.5~GeV$|$ $<$ 40~GeV and $|m_{l_{c}\nu}^{reco}$ - 80.4~GeV$|$ $<$ 30~GeV.

\section{Signal and control regions}

The agreement between data and simulated samples of the expected background is investigated through the control regions distributions and their inclusion in the fit. With that, the rescalling of the background expectations to the best fit with observed data and the reduction of the background uncertainties are applied. Therefore, five distinct control regions were defined focusing on $t\bar{t}Z$, $WZ$, $ZZ$ processes and non-prompt leptons backgrounds and distinguishing from the signal region with the number of jets, $b$-tagged jets, leptons or the presence of a $Z$ boson candidate.

\begin{figure}[htb]
\centering
\begin{subfigure}[b]{0.35\textwidth}
\includegraphics[width=\textwidth]{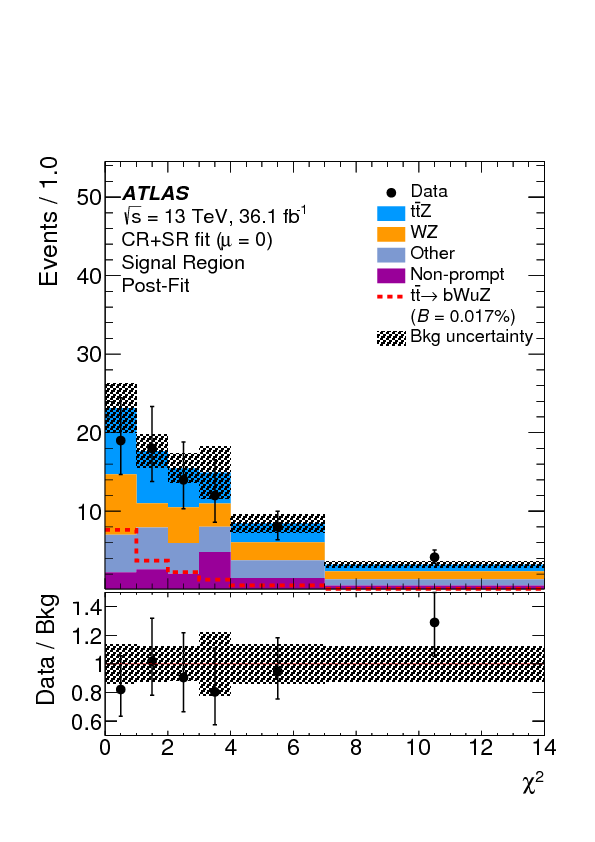}
\caption{}
\end{subfigure}\qquad
\begin{subfigure}[b]{0.35\textwidth}
\includegraphics[width=\textwidth]{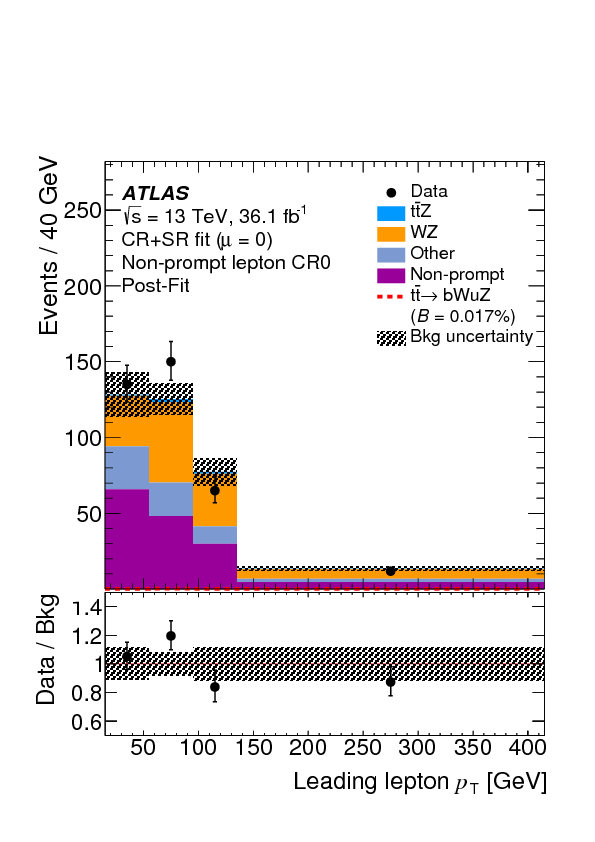}
\caption{}
\end{subfigure}\\
\begin{subfigure}[b]{0.35\textwidth}
\includegraphics[width=\textwidth]{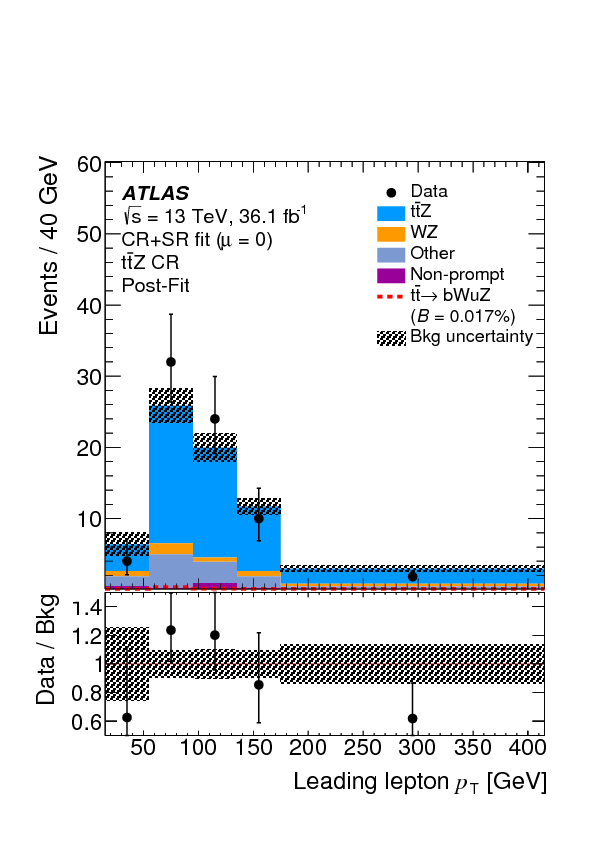}
\caption{}
\end{subfigure}\qquad
\begin{subfigure}[b]{0.35\textwidth}
\includegraphics[width=\textwidth]{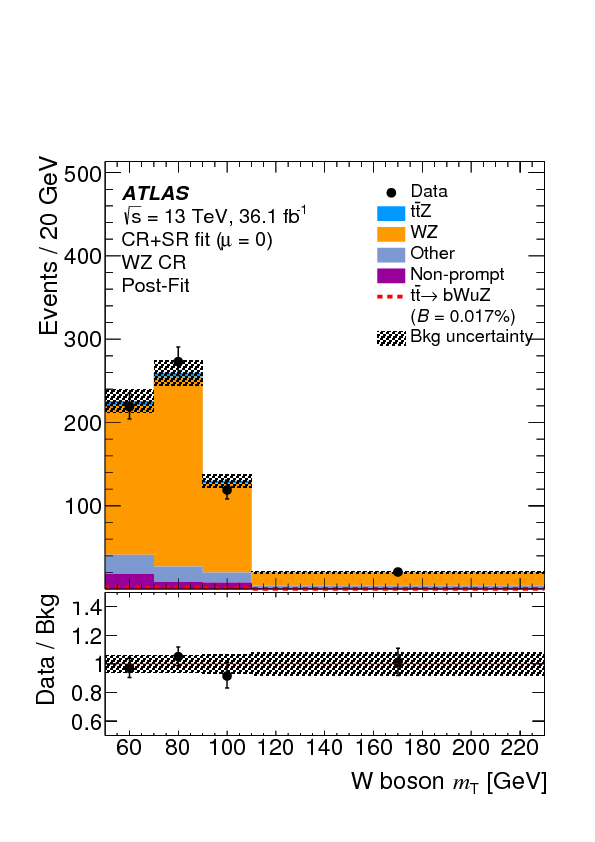}
\caption{}
\end{subfigure}
\caption{Expected (filled histogram) and observed (points with error bars) distributions after the combined fit under the background-only hypothesis of (a) the $\chi^2$ of the kinematical reconstruction in the signal region; (b) $p_{\rm T}$ of the leading lepton in the non-prompt lepton control region with the $b$-tagged jet veto; (c) $p_{\rm T}$ of the leading lepton in the $t\bar{t}Z$ control region; (d) the transverse mass of the $W$ boson in the $WZ$ control region. For comparison, distributions for the FCNC $t\bar{t}$ $\to$ $bWuZ$ signal (dashed line), normalized to the observed limit, are also shown. The dashed area represents the total uncertainty in the background prediction.}
\label{fig:kinematic}
\end{figure}

Kinematical distributions from the signal region and three of the control regions are represented in Figure~\ref{fig:kinematic}. For the signal region, the $\chi^2$ distribution demonstrates a lower value of this variable for the signal when compared with the background and the distinct contributions from the main backgrounds. The remaining distributions present a proper isolation of the dedicated backgrounds with a good agreement between data and MC simulation. 

\section{Results}

 To extract the FCNC signal, a binned likelihood fit is performed to different variable distributions simultaneously in the signal and control regions. 
 In the signal region, the $\chi^2$ variable from the kinematical reconstruction of the top-quark is used for the signal-background discrimination. Additional distributions from the control regions were added to the fit and allowed a tighter constraint of background normalisations and systematic uncertainties compared with a fit where just the signal region is used.
 A good agreement between data and expectation from the background-only hypothesis is observed and no evidence of a FCNC signal is found. The CL$_{s}$ method \cite{CLs,CLs2} assuming that only one FCNC mode contributes is used to obtain the upper limits on the correspondent branching ratios. The 95\% confidence level observed (expected) upper limits on the $t\to qZ$ branching ratio are $\mathcal{B}$($t\to uZ$) $<$ 1.7 $\times$ 10$^{-4}$ (2.4 $\times$ 10$^{-4}$) and $\mathcal{B}$($t\to cZ$) $<$ 2.4 $\times$ 10$^{-4}$ (3.2 $\times$ 10$^{-4}$) \cite{Results}. These limits constitute the most stringent limits to date being a factor of around two better than the ones obtained by the CMS Collaboration at the same centre-of-mass energy and integrated luminosity \cite{CMS}. The observed limits on branching ratio were also converted into a constraint on the values of effective field theory operators contributing to the FCNC decays of the top-quark \cite{TopFCNC}.




\end{document}